\documentclass[12pt]{article}
\usepackage{fancyhdr}

\usepackage{mathrsfs}
\usepackage[T1]{fontenc}
\usepackage{mathpazo}
\usepackage{setspace}
\usepackage{amsfonts}
\usepackage{amssymb}
\usepackage{amsmath}
\usepackage{epsfig}
\usepackage{latexsym}
\usepackage{color}
\usepackage{graphicx}
\usepackage{nicefrac}
\usepackage[latin1]{inputenc}
\usepackage{slashed}
\usepackage{multirow}
\usepackage{comment}
\usepackage{soul}
\usepackage{hyperref}
\usepackage[nosort]{cite}
\usepackage{datetime}
\usepackage{braket}

\usepackage[titletoc]{appendix}

\def\hybrid{\topmargin -20pt    \oddsidemargin 0pt
	\headheight 0pt \headsep 0pt
	\textwidth 6.25in       
	\textheight 9.25in       
	\marginparwidth .875in
	\parskip 5pt plus 1pt   \jot = 1.5ex}

\hybrid
\begin{document}
\numberwithin{equation}{section}
\begin{titlepage}
	\begin{center}
\vskip .5 in

{\large\bf Fermionic path integrals and correlation dynamics in a 1D XY system}

\vskip 0.35in

{\bf I. Lyris}, {\bf P. Lykourgias}, {\bf A. I. Karanikas}
\vskip 0.1in

{\em{}Department of Nuclear and Particle Physics,\\ Faculty of Physics, National and Kapodistrian University of Athens,\\15784 Athens, Greece
}

\vskip 0.1in

\vskip 0.1in

{\footnotesize \texttt giannislyris@phys.uoa.gr, panoslyk@phys.uoa.gr, akaran@phys.uoa.gr}

\vskip .43in
\end{center}

\centerline{\bf Abstract}
\noindent
We derive time dependent correlation functions in an one dimensional XY spin model
with the use of generating functionals, the latter being defined as path integrals over fermionic
coherent states. We focus on the proper construction of the aforementioned integrals in order to
avoid the inconsistencies usually encountered in the literature. The static limit of our results successfully reproduces the known ones, confirming the validity of our construction and allowing for further investigation of the dynamics.
In the same context, we examine the case of a general driven transverse magnetic field, for which case we derive formulas for the equal-time correlation functions and 
confirm the consistency of our results with the Kibble-Zurek mechanism.

\vskip .4in
\noindent
\end{titlepage}
\vfill
\eject
\tableofcontents

\def\baselinestretch{1.2}
\baselineskip 20 pt
\section{Introduction} 
The Feynman path integral formalism is the most powerful tool for taking into account quantum behaviour via classical computations \cite{1,2}. Ideally suited for semiclassical
calculations, the path integral machinery provides a variety of analytical
methods for studying the dynamics of quantum correlations in closed and open quantum systems \cite{1,3}. The extension of path integration to the ordinary complex
plane $\mathbb{C}$ through the Glauber coherent states \cite{4}, to the complex non-flat manifold
$\bar{\mathbb{C}}$ through the $su(2)$ spin coherent states \cite{5,6,7}, and to fermionic systems
through the fermionic coherent states \cite{1,8} has allowed for the application of path integral techniques to the study of many-body systems \cite{9,10}. These systems are of
great interest for both the condensed matter physics and the quantum information science due to the fact that they naturally support entangled states. Correlations in these states have a fundamental quantum character as they do not have a classical counterpart, and can serve both as the means for understanding quantum phase
transitions and as the main tool for quantum information processing \cite{11,12,13}.
During the last years, there have been considerable advances in the study of the static properties
and the dynamics of closed many-body quantum systems both at experimental and
theoretical level \cite{14,15}. However, despite these advances, the usage of path integral techniques in the corresponding analysis is rather restricted. The main reason is that, currently, there is no universally accepted way to define the path integral on complexified spaces, spanned by the coherent state bases, for a general system written in terms of bosonic, spin or fermionic operators \cite{16,17,18,19}. When free from conceptual and
defining issues, path integrals over coherent states can provide a wide palette of
techniques, analytical and numerical, for the analysis of quantum systems either
closed or open. In the current paper we make a step towards this direction by developing an inconsistency-free method for the study of spin systems
through the use of fermionic coherent state path integrals. 
\newline
\newline
In the first part of our study we discuss the construction of path integrals corresponding to systems written in terms of fermionic creation and annihilation operators. Due to the anticommuting character of such operators, the variables pertaining the path integral need to anticommute as well, which means they are of Grassmann nature. As we will thoroughly discuss, while the construction of such path integrals is considered well known \cite{1}, inconsistencies similar to those in the cases of bosonic and spin systems, appear here as well. To overcome these we develop a definite procedure, based on the recently proposed path integral representation of Majorana fermions \cite{26}, which as we show leads to inconsistency-free fermionic coherent state path integral expressions capable of interpreting fermionic quantum systems. In the second part of the present study we analyse the dynamics of the XY model in one dimension, which is a model perfectly suited for applying the proposed formalism, since it is exactly solvable
and yet has a quite rich structure that supports a quantum phase transition \cite{11,12,20,21}. For this system, by taking advantage of the path integral representation, we calculate the exact time-dependence of the ground state correlation functions, while as a check we also reproduce the results known for the static case, pertaining to the entanglement entropy. We then consider the case of a general driven transverse magnetic field, for which we showcase that path integration methods allow for the computation of time-dependent equal-time correlation functions in the form of convergent series. Finally, we confirm that the
Kibble-Zurek mechanism characterizes the passage of a driven system through a quantum critical point \cite{22,23}.
\newline
\newline
The structure of the paper is the following: In Section $2.1$ we present the issues
arising in the construction of path integrals over fermionic coherent states and 
examine the solution that already has been proposed for the confrontation of similar
problems appearing in integrals over bosonic coherent states \cite{24,25}. In Section $2.2$
we discuss the recently proposed path integral representation for Majorana fermions
\cite{26} proving that it can be used as an intermediate step towards an inconsistency-free
path integral description of fermionic systems. In Section $3.1$ we define the path integral representation of the 1D XY model in order to derive the time-dependent ground state correlators needed for the calculation of the entanglement entropy and we compare with the known static result. In Section $3.2$ we examine the case of a general driven transverse magnetic field and investigate the time dependence of the ground state correlators we are interested in. We also find that the driving of the system through the critical point is consistent with the Kibble-Zurek scaling mechanism. Finally, in Section $4$ we present
our conclusions and the perspectives of our work. The paper is accompanied by two appendices. In Appendix A we present the path integral calculation of the partition function pertaining to the toy-model $\hat{H}=-\omega\hat{\vec{S}}_1\cdot\hat{\vec{S}}_2$. In Appendix B we present details of the calculations pertaining to the results that appear in Section $3$.
\section{Functional integration over Grassmann variables}
\subsection{Fermionic coherent state path integrals}
In this section we shall discuss path integration in the space spanned by the fermionic coherent states, that is, the eigenstates of the fermionic annihilation operator $\hat{\psi}$, the eigenvalues of which are complex Grassmann variables
\begin{equation}
\hat{\psi}\ket{\zeta}=\zeta\ket{\zeta}.
\end{equation}
These states form an overcomplete basis 
\begin{equation}
\int\frac{d\bar{\zeta}d\zeta}{\pi}\ket{\zeta}\bra{\zeta}=\ket{0}\bra{0}+\ket{1}\bra{1}={\mathbb{1}}, \label{compl}
\end{equation}
which can be used to transcribe fermionic amplitudes into the context of path integrals \cite{1}. The need for a careful reconsideration of this transcription can be traced back in the issues occurring in both bosonic and spin coherent state path integrals, issues that question even their very meaning \cite{16}. By referring to the Jordan-Wigner \cite{27} transformation one expects similar problems to occur when path integration is performed in the basis of fermionic coherent states. In the present work we shall examine correlations in spin systems using their corresponding fermionic coherent state path integral representation, thus the use of a formulation free of inconsistencies is of vital importance.
\newline
\newline
To begin with, consider the partition function of a system, the Hamiltonian of which is expressed in terms of fermionic creation and annihilation operators:
\begin{equation}
Z=\text{Tr}e^{-\beta\hat{H}}=\int\frac{d\bar{\zeta}d\zeta}{\pi}\braket{-\zeta|e^{-\beta\hat{H}}|\zeta}.
\end{equation}
By dividing $\beta$ in $N+1=\beta/\epsilon$ segments and inserting the completeness relation (\ref{compl}) in each division, the following formal result is achieved \cite{1} at the limit $\epsilon\rightarrow0$:
\begin{equation}
Z=\int_{AP}\mathcal{D}\bar{\zeta}\mathcal{D}\zeta{\text{exp}\left\{-\int_{-\beta/2}^{\beta/2}d\tau\left[\bar{\zeta}\dot{\zeta}+H\left(\bar{\zeta},\zeta\right)\right]\right\}}. \label{partition}
\end{equation}
The last expression is formal in the sense that both the form of the function $H$ that represents the quantum Hamiltonian $\hat{H}$, and its discrete ancestor, must be carefully defined in order to avoid any inconsistencies.
To give a very simple example, consider the case of the fermionic oscillator 
\begin{equation}
\hat{H}=\omega\left(\hat{\psi}^\dagger\hat{\psi}-\frac{1}{2}\right) \label{fermosc}
\end{equation}
which is connected to the spin Hamiltonian $\hat{H}=-\omega\hat{S}_z$ via the Jordan-Wigner transformation. The partition function in this case can be trivially computed without any reference to path integration:
\begin{equation}
Tre^{-\beta\hat{H}}=e^{\omega\beta/2}+e^{-\omega\beta/2}=2\text{cosh}\left(\omega\beta/2\right).
\end{equation}
Trying to obtain the same result by means of path integration, we follow the standard discretization procedure and adopt for the classical function $H$ the expression 
\begin{equation}
\frac{H(\zeta_n,\zeta_{n-1})}{\braket{\zeta_n|\zeta_{n-1}}}=\omega\left(\bar{\zeta}_n\zeta_{n-1}-\frac{1}{2}\right)\underset{N\rightarrow\infty}{\longrightarrow}\omega\left(\bar{\zeta}\zeta-\frac{1}{2}\right). \label{limit}
\end{equation}
Thus, for the partition function we get the following result
\begin{equation}
Z=e^{\omega\beta/2}\int_{AP}\mathcal{D}\bar{\zeta}\mathcal{D}\zeta{\text{exp}\left\{-\int_{-\beta/2}^{\beta/2}d\tau\bar{\zeta}\left(\partial_\tau+\omega\right)\zeta\right\}}. \label{simple}
\end{equation} 
The integral involved in the above expression can be evaluated using the formula \cite{1}:
\begin{equation}
\text{Tr}\text{ln}(\partial_\tau+\omega)=\beta\int_{0}^{\omega}d\omega'G_{\omega'}(\tau,\tau) \label{formula}
\end{equation}
where $G_{\omega'}(\tau,\tau')$ is the Green's function satisfying $(\partial_\tau+\omega)G_{\omega}(\tau,\tau')=\delta(\tau-\tau')$ with antiperiodic boundary conditions $G_{\omega}\left(-\beta/2,\tau\right)=-G_{\omega}\left(\beta/2,\tau\right)$
\begin{equation}
G_\omega(\tau,\tau')=\left[\Theta(\tau-\tau')-\left(1+e^{\beta\omega}\right)^{-1}\right]e^{-\omega(\tau-\tau')}. \label{retard}
\end{equation}
The function $\Theta(\tau-\tau')$ is usually chosen \cite{1} to be the Heaviside function for which $\Theta(0)=\frac{1}{2}$. This choice is ultimately related to the requirement for path integrals to be defined through dimensional regularization \cite{1}, which in turn gives a definite prescription for $G(0)$. However, due to the unavoidable discontinuity of the Green's function at $\tau=\tau'$, questions arise \cite{24,25} regarding the meaning of $G(\tau,\tau)$ in Eq. (\ref{formula}), since different prescriptions yield different results. Indeed, by adopting the symmetric prescription indicated in Eq. (\ref{retard}), we find for the partition function the wrong result
\begin{equation}
Z=2e^{\omega\beta/2}\text{cosh}\left(\omega\beta/2\right). \label{wrong}
\end{equation}
For the cure of this awkward situation, it has been proposed \cite{24,25} to take into account the discrete ancestor of the continuous Hamiltonian (see Eq. (\ref{limit})) and to use the limit form $G_\omega(\tau,\tau)=\left(1+e^{\beta\omega}\right)^{-1}$ in Eq. (\ref{formula}). In this way, the path integral of Eq. (\ref{simple}) yields the following result
\begin{equation}
\int_{AP}\mathcal{D}\bar{\zeta}\mathcal{D}\zeta{\text{exp}\left\{-\int_{-\beta/2}^{\beta/2}d\tau\bar{\zeta}\left(\partial_\tau+\omega\right)\zeta\right\}}=1+e^{-\beta\omega}. \label{correct}
\end{equation}
Thus, the correct answer for the partition function is recovered. Nevertheless, in the case of path integration in terms of bosonic and spin coherent states, this prescription is not enough \cite{16,17,18,19} for curing the inconsistencies appearing in less trivial systems, such as the Bose-Hubbard model, or the spin system $\hat{H}=\omega\hat{S}_z^2$, with $s>\frac{1}{2}$. In the fermionic case, the Grassmannian character of the fields does not permit non-linearities to appear, and thus the asymmetric prescription $G(\tau,\tau+0)$ can successfuly cope with the calculation of the path integral in that specific case. However, this prescription is strongly tied to the asymmetric form of the underlying discrete action, an action that makes the discrete path integral not invariant under transformations that should leave the continuous path integral intact. For example, the discrete ancestor of the continuous path integral is not invariant under canonical transformations, a fact that contradicts the physical demand for the path integral to share this invariance with classical mechanics \cite{1}.
Although these observations seem to be of academic nature regarding the calculation of the partition function, we shall confirm that they turn out to be quite important when the calculation of correlation functions becomes the main issue. 
\subsection{The Majorana path integral}

To face the inconsictencies appearing in the standard formulation of path integration of bosonic and spin systems in the presence of interactions, we have proposed \cite{17,19} a simple approach which circumvents the direct construction, through the introduction of Hermitian ''position'' and ''momentum'' operators. The construction of the Feynman phase space path integral is then possible with the use of the previous operators' eigenvalues. This integral can be then transcribed to the desired complex, flat or non-flat manifold, through a canonical transformation. An extension of such an approach though is not possible for the fermionic case, as the corresponding Hermitian operators are Majorana fermions, for which the canonical quantization is highly non-trivial and the formalism of coherent states does not exist. The process of identifying the correct continuum limit thus should proceed differently. 
\newline
\newline
Recently \cite{26}, it has been proved that the quantization of a Majorana system through the path integral representation is possible. In the relevant construction, the Majorana system is considered as a constrained system that is handled via the Faddeev-Jackiw method \cite{28}. In this approach one can directly define the path integral pertaining to the system at hand, after obtaining the classical action which is consistent with the corresponding quantum theory. It is on this approach that we will rely to specify the structure of path integrals over fermionic coherent states, for systems like the one in Eq. (\ref{fermosc}). To be concrete, let us examine again the trivial case of the simple fermionic oscillator ($\ref{fermosc})$. By introducing the Majorana operators 
\begin{equation}
\hat{\psi}+\hat{\psi}^\dagger=\hat{\gamma}_1,\quad\hat{\psi}^\dagger-\hat{\psi}=-i\hat{\gamma}_2;\quad\{\hat{\gamma}_a,\hat{\gamma}_b\}=\delta_{ab},
\end{equation}
the Hamiltonian (\ref{fermosc}) assumes the form $\hat{H}_{cl}=-i\frac{\omega}{2}\hat{\gamma}_2\hat{\gamma}_1$. The construction of the corresponding path integral proceeds then via the Faddeev-Jackiw method and dictates \cite{26} the form $H_{cl}=-i\frac{\omega}{2}\gamma_2\gamma_1$ for the classical function which weighs the path integration, with $\{\gamma_a,\gamma_b\}=0$. The integral constructed in this way represents the partition function of the system as a path integral over real Majorana Grassmann variables. It is then an inevitable demand for this integral to be connected with the corresponding integral over the complex Grassmann variables, through the canonical transformation $\gamma_1=\zeta+\bar{\zeta},$ $-{i}\gamma_2=\bar{\zeta}-\zeta$. This approach yields the Hamiltonian $H_M=\omega\bar{\zeta}\zeta$ as the proper weight for the integration over fermionic paths 
\begin{equation}
Z=\int_{AP}\mathcal{D}\bar{\zeta}\mathcal{D}\zeta{\text{exp}\left\{-\int_{-\beta/2}^{\beta/2}d\tau\bar{\zeta}\left(\partial_\tau+\omega\right)\zeta\right\}}.
\end{equation} 
It is worth noting that for the above-mentioned canonical transformation to be valid, the discretization prescription underlying the continuous form must be the symmetric one $\bar{\zeta}_n\zeta_n\underset{N\rightarrow\infty}{\longrightarrow}\bar{\zeta}\zeta$. Thus, for the calculation of the integral we must use for the Green's function (\ref{retard}) the symmetric limit value $G_\omega(0)=\frac{1}{2}-\left(1+e^{\beta\omega}\right)^{-1}$. In this way, the correct result is produced.
\newline\newline
The calculation we presented can be summarized in a simple proposal: To use the path integral formalism for a system, the quantum Hamiltonian of which is given in terms of fermionic creation and annihilation operators, begin by rewriting it in terms of Majorana operators and continue by replacing these with the corresponding real Grassmann variables according to the Faddeev-Jackiw procedure. Next, perform a canonical change of variables to get the classical Hamiltonian that must weigh the paths over fermionic coherent states. This whole construction fixes the discrete ancestor of the continuous expressions to be the symmetric one.\newline\newline
To demonstrate the general form of the quantum Hamiltonians we are going to deal with, consider the spin Hamiltonian 
\begin{equation}
\hat{H}=-\sum_{j=1}^{N}\left[a_j\sigma_j^x\sigma_{j+1}^x+b_j\sigma_j^y\sigma_{j+1}^y+c_j\sigma_j^z\sigma_{j+1}^z+h_j\sigma_j^z\right]. \label{spinh}
\end{equation}
Applying the Jordan-Wigner transformation \cite{28.1}
\begin{align}
\begin{split}
&\sigma_j^x=\left[\prod_{k=1}^{j-1}\left(1-2\hat{\psi}^\dagger_k\hat{\psi}_k\right)\right]\left(\hat{\psi}_j^\dagger+\hat{\psi_j}\right),\\
&\sigma_j^y=i\left[\prod_{k=1}^{j-1}\left(1-2\hat{\psi}^\dagger_k\hat{\psi}_k\right)\right]\left(\hat{\psi}_j^\dagger-\hat{\psi_j}\right),\\
&\sigma_j^z=1-2\hat{\psi}^\dagger_j\hat{\psi}_j, \label{jord}
\end{split}
\end{align}
we can re-express (\ref{spinh}) in terms of fermionic creation and annihilation operators as
\begin{align}
\hat{H}=&-\sum_{j=1}^{N}\bigg[a_j\left(\hat{\psi}^\dagger_{j}-\hat{\psi}_{j}\right)\left(\hat{\psi}^\dagger_{j+1}+\hat{\psi}_{j+1}\right)+b_j\left(\hat{\psi}^\dagger_{j+1}-\hat{\psi}_{j+1}\right)\left(\hat{\psi}^\dagger_{j}+\hat{\psi}_{j}\right)\nonumber+\\
&+c_j\left(1-2\hat{\psi}^\dagger_j\hat{\psi}_j\right)\left(1-2\hat{\psi}^\dagger_{j+1}\hat{\psi}_{j+1}\right)+h_j\left(1-2\hat{\psi}^\dagger_j\hat{\psi}_j\right)\bigg]. \label{fermionh}
\end{align}
Introducing the Majorana operators 
\begin{equation}
\hat{\gamma}_{2j-1}=\hat{\psi}^\dagger_j+\hat{\psi}_j,\quad-i\hat{\gamma}_{2j}=\hat{\psi}^\dagger_j-\hat{\psi}_j;\quad\left\{\hat{\gamma}_{j},\hat{\gamma}_{k}\right\}=\delta_{jk}, \label{majora}
\end{equation}
the Hamiltonian (\ref{spinh}) assumes the form 
\begin{equation}
\hat{H}=i\sum_{j=1}^{N}\left(a_j\hat{\gamma}_{2j}\hat{\gamma}_{2j+1}+b_j\hat{\gamma}_{2j+2}\hat{\gamma}_{2j-1}+ic_j\hat{\gamma}_{2j-1}\hat{\gamma}_{2j}\hat{\gamma}_{2j+1}\hat{\gamma}_{2j+2}+h_j\hat{\gamma}_{2j-1}\hat{\gamma}_{2j}\right). \label{majoranh}
\end{equation}
Using the Faddeev-Jackiw procedure we can now identify the classical Hamiltonian weighing the functional integral over Majorana variables by replacing the Majorana operators with classical real Grassmann variables $\gamma_j$.  By changing these back to the complex Grassmann variables via the canonical transformation
\begin{equation}
{\gamma}_{2j-1}=\bar{\zeta}_j+\zeta_j,\quad-i{\gamma}_{2j}=\bar{\zeta}_j-\zeta_j;\quad\left\{\zeta_{j},{\zeta}_{k}\right\}=\left\{\bar{\zeta}_{j},{\zeta}_{k}\right\}=\left\{\bar{\zeta}_{j},\bar{\zeta}_{k}\right\}=0, \label{majoravar}
\end{equation}
we get for the Hamiltonian weighing the functional integral over complex Grassmann variables
\begin{align}
{H}_{cl}=&\sum_{j=1}^{N}\bigg[a_j\left(\bar{\zeta}_{j+1}+\zeta_{j+1}\right)\left(\bar{\zeta}_{j}-\zeta_{j}\right)+b_j\left(\bar{\zeta}_{j}+\zeta_{j}\right)\left(\bar{\zeta}_{j+1}-\zeta_{j+1}\right)\nonumber-\\
&-4c_j|\zeta_j|^2|\zeta_{j+1}|^2+2h_j|\zeta_j|^2\bigg], \label{finalh}
\end{align}
which is also the Hamiltonian weighing the corresponding fermionic coherent state path integrals, as the one in (\ref{partition}). It is worth noting that the consistent definition of the path integral representation of transition amplitudes in the continuum limit is not only an academic issue, as it makes available a variety of techniques - borrowed from the quantum field theory toolkit - to support the study of systems of interest in the fields of condensed matter physics and quantum information science. The formalism is, for example, ideally suited for the calculation of time dependent correlation functions, either exactly or semiclassicaly. In the present paper, we focus on the dynamics of ground state correlators in a spin-chain system described by the $XY$ model. Besides the exact evaluation of correlators' time-dependence, we shall recover the correct equal time results that have been evaluated by different means, in order to perform a series of non-trivial checks regarding our procedure. As a concrete example we present, in Appendix A, a simple calculation pertaining to the two-spin system $\hat{H}=-\omega\hat{\vec{S}}_1\cdot\hat{\vec{S}}_2$.
\section{Correlator dynamics in the XY model}
\subsection{Time dependent correlations}
The study of entanglement in 1D, spin-1/2 chain models, is of great interest not only in the field of condensed matter physics, but also in quantum information science, where entangled states are of fundamental importance in information processing. The $XY$ model is a well-known and exactly solvable model that excibits a quantum phase transition. This transition signals the onset of long-range correlations in the ground state of the system, and it is of purely quantum mechanical nature,  as it is connected to the entanglement properties of the ground state \cite{11,12,13}. Thus, the $XY$ model constitutes the ideal stage for the application of the path integral formalism, since it provides the tools to explore the dynamics of the vacuum correlation functions.
\newline
\newline
The couplings in the anisotropic $XY$ model are defined \cite{11,12,13} to be $a_j=(1+r)/2$, $b_j=(1-r)/2$ and $h_j=h$, $c_j=0$ $\forall{j}$. Thus, the Hamiltonian (\ref{finalh}) reads as follows 
\begin{equation}
H_{XYcl}=\sum_{j=1}^{N}\left[r\left(\zeta_j\zeta_{j+1}-\bar{\zeta}_j\bar{\zeta}_{j+1}\right)+\left(\zeta_j\bar{\zeta}_{j+1}-\bar{\zeta}_j\zeta_{j+1}\right)+2h\bar{\zeta}_j\zeta_j\right]. \label{xyham}
\end{equation}
For the purposes of this chapter we introduce Grassmann sources, redefining the generating functional as
\begin{equation}
Z[J]=\int_{AP}\mathcal{D}\bar{\zeta}\mathcal{D}\zeta{}\text{exp}\left\{-\int_{-\frac{\beta}{2}}^{\frac{\beta}{2}}d\tau\left[\sum_{j=1}^{N}\bar{\zeta}_j\dot{\zeta}_j+H_{XYcl}-i\sum_{j=1}^{N}(\bar{J}_j\zeta_j+\bar{\zeta}_jJ_j)\right]\right\}. \label{genfun}
\end{equation}
The functional derivatives of this integral generate the expectation values of operators as
\begin{eqnarray}
\frac{\delta^2\text{ln}Z[J]}{\delta{J}_b(\tau_2)\delta\bar{J}_a(\tau_1)}\bigg|_{J=0}=\braket{\hat{T}\left(\hat{\psi}^\dagger_b(\tau_2)\hat{{\psi}}_a(\tau_1)\right)}_c.
\end{eqnarray}
Here $\hat{T}$ signifies the time ordering of the operators, which is implied by the path integral procedure, and the index $c$ the connected part of the expectation values. This result is true only if at least $b\neq{a}$ or $\tau_2\neq\tau_1$, since when all indices (time and site) are equal one should expect similar issues as those in the identification of the correct Hamiltonian symbol. To identify the correct way to handle such cases we will now see that it is enough to understand how the simple spin operators of the original system are mapped in this context.
\newline
\newline
Spin-spin correlators of the form $\braket{\sigma_i^\alpha\sigma_j^\beta}_c=\braket{\sigma_i^\alpha\sigma_j^\beta}-\braket{\sigma_i^\alpha}\braket{\sigma_j^\beta},(\alpha,\beta=x,y,z)$ are physically quite important as they probe the entanglement content of the ground state \cite{11}. Such types of correlators can be produced by applying on the generating integral (\ref{genfun}) the appropriate functional derivatives: 
\begin{equation}
\braket{\hat{T}\left(\sigma_i^\alpha(\tau_2)\sigma_j^\beta(\tau_1)\right)}_c=\Delta_i^\alpha(\tau_2)\Delta_j^\beta(\tau_1)\text{ln}Z[J]|_{J=0}. \label{spincor}
\end{equation}
To determine the form of these operations we can begin from the defining relation
\begin{equation}
\braket{\sigma_j^\alpha}=\frac{1}{Z(\beta)}\text{Tr}\left(e^{-\beta\hat{H}(\hat{\psi}^\dagger,\hat{\psi})}\sigma_j^\alpha(\hat{\psi}^\dagger,\hat{\psi})\right), \label{sigmaz}
\end{equation}
where the operators $\sigma_j^\alpha(\hat{\psi}^\dagger,\hat{\psi})$ are defined through the Jordan-Wigner transformation (\ref{jord}). The next step is to interpret Eq. (\ref{sigmaz}) as a path integral over fermionic coherent states. In the standard formulation, presented in section $2.1$, the operator $\sigma^z_j=1-2\hat{\psi}^\dagger_j\hat{\psi}_j$ is interpreted by the classical function $1-2\bar{\zeta}_j\zeta_j$ (see Eq. (\ref{limit})), while the Faddeev-Jackiw approach, presented in Section 2.2, yields the function $-2\bar{\zeta}_j\zeta_j$. Since the $\sigma^{z}_{j}$ operator enters non-trivially in each and every spin correlation function, it is evident that the two prescriptions produce different results. As the static correlation functions for the XY model are known, they can serve as a criterion for distinguishing between the two approaches. In the following we argue that the correct results are produced through the Faddeev-Jackiw construction which yields the results
\begin{align}
\begin{split}
&\braket{\sigma_j^x}=\frac{1}{Z(\beta)}\underset{AP}{\int}\mathcal{D}\bar{\zeta}\mathcal{D}\zeta{e}^{-S(\bar{\zeta},\zeta)}\left(\prod_{k=1}^{j-1}2\zeta_k\bar{\zeta}_k\right)(\bar{\zeta}_j+\zeta_j),\\
&\braket{\sigma_j^y}=\frac{i}{Z(\beta)}\underset{AP}{\int}\mathcal{D}\bar{\zeta}\mathcal{D}\zeta{e}^{-S(\bar{\zeta},\zeta)}\left(\prod_{k=1}^{j-1}2\zeta_k\bar{\zeta}_k\right)(\bar{\zeta}_j-\zeta_j),\\
&\braket{\sigma_j^z}=\frac{2}{Z(\beta)}\underset{AP}{\int}\mathcal{D}\bar{\zeta}\mathcal{D}\zeta{e}^{-S(\bar{\zeta},\zeta)}\zeta_j\bar{\zeta}_j. \label{corel}
\end{split}
\end{align}
Thus, the operations in Eq. (\ref{spincor}) are defined as
\begin{align}
\begin{split}
&\Delta_j^x(\tau)\equiv\left(\prod_{k=1}^{j-1}2\frac{\delta^2}{\delta\bar{J}_k(\tau)\delta{J}_k(\tau)}\right)\left(\frac{\delta}{\delta{J}_j(\tau)}+\frac{\delta}{\delta{}\bar{J}_j(\tau)}\right),\\
&\Delta_j^y(\tau)\equiv{i}\left(\prod_{k=1}^{j-1}2\frac{\delta^2}{\delta\bar{J}_k(\tau)\delta{J}_k(\tau)}\right)\left(\frac{\delta}{\delta{J}_j(\tau)}-\frac{\delta}{\delta{}\bar{J}_j(\tau)}\right),\\
&\Delta_j^z(\tau)\equiv2\frac{\delta^2}{\delta\bar{J}_j(\tau)\delta{J}_j(\tau)}. \label{operat}
\end{split}
\end{align}
If the ground state of the system is unique Eq. (\ref{spincor}) produces, at the limit $\beta\rightarrow\infty$, the ground state expectation value of the operator. In the case of degeneracy, the zero-temperature limit projects on an equiprobable mixture of the degenerate ground states. At the limit $\beta\rightarrow\infty$ and after the Wick rotation $\tau\rightarrow{it}$ the operation (\ref{spincor}) generates the time-dependent two-point vacuum expectation value $\braket{\hat{T}\left(\sigma^a_i(t_2)\sigma^b_j(t_1)\right)}_c$ \cite{28.5}. This result also indicates the correct prescription for the equal time fermionic correlation functions\footnote{In the case that both operators are holomorphic or antiholomorphic no extra contribution appears at the equal time limit.}, as 
\begin{eqnarray}
\braket{\hat{T}\left(\hat{\psi}^\dagger_b(\tau)\hat{\psi}_a(\tau)\right)}_c=\frac{\delta^2\text{ln}Z[J]}{\delta{J}_b(\tau)\delta{\bar{J}}_a(\tau)}\bigg|_{J=0}+\frac{1}{2}\delta_{ab} \label{eq}
\end{eqnarray}
The study of the static entanglement entropy has been based \cite{11,12} on the equal time version of the above defined correlation functions. In the current section, we shall use the aforementioned path integral technique to investigate the dynamics of ground-state correlation functions, and as a result of the underlying entanglement, while in the static case we will also confirm the known static results. To deal with the path integration weighted by (\ref{xyham}), we shall follow the usual \cite{11,12,13,29} tactic, which is based on the introduction of the Fourier transforms
\begin{equation}
\zeta_i=\frac{1}{\sqrt{N}}\sum_{m=0}^{N-1}e^{i\frac{2\pi}{N}\left(m+\frac{1}{2}\right)j}c_m\quad{J}_i=\frac{1}{\sqrt{N}}\sum_{m=0}^{N-1}e^{i\frac{2\pi}{N}\left(m+\frac{1}{2}\right)j}\lambda_m. \label{Fourier}
\end{equation}
Note, at this point, that transformation (\ref{Fourier}) implies the anti-periodic condition $\zeta_{N+1}=-\zeta_1$. In turn, this is connected to a chain of even number of fermions. If this number is odd, the change $m+1/2\rightarrow{m}$ is required \cite{29}. Inserting (\ref{Fourier}) in (\ref{xyham}) we are able to write the classical Hamiltonian in the form
\begin{equation}
H_{XYMcl}=\sum_{m=0}^{N/2-1}H_m,\quad{H}_m=2\left(\begin{array}{cc}\bar{c}_m & c_{N-m-1} \end{array} \right)\left(\begin{array}{cc}k_m & -il_m \\il_m & -k_m \end{array} \right)\left(\begin{array}{cc}c_m \\\bar{c}_{N-m-1}\end{array} \right), \label{abbreviation}
\end{equation}
where we have introduced the abbreviations
\begin{equation}
k_m=h-\text{cos}\frac{2\pi}{N}\left(m+\frac{1}{2}\right)\quad{}l_m=r\text{sin}\frac{2\pi}{N}\left(m+\frac{1}{2}\right). \label{abbreviations}
\end{equation}
The crucial observation here is that the interactions connect only the fields $\bar{c}_m$ with $c_m$ or $\bar{c}_{N-m-1}$, and the fields $c_m$ with $\bar{c}_m$ or $c_{N-m-1}$. Thus, the generating functional can be factorized as
\begin{equation}
Z[J]=\prod_{m=0}^{N/2-1}Z_m[J],\quad{Z}_m[J]=\int_{AP}\mathcal{D}\bar{c}_m\mathcal{D}c_m\mathcal{D}\bar{c}_{N-m-1}\mathcal{D}c_{N-m-1}e^{-S_m[J]}, \label{observation}
\end{equation}
where
\begin{align}
S_m[J]=&\int_{-\infty}^{+\infty}d\tau\bigg\{\left(\bar{c}_m\dot{c}_{m}+\bar{c}_{N-m-1}\dot{c}_{N-m-1}\right)+H_m-\nonumber\\
&-i\left(\bar{\lambda}_mc_m+\bar{c}_m\lambda_m+\bar{\lambda}_{N-m-1}c_{N-m-1}+\bar{c}_{N-m-1}\lambda_{N-m-1}\right)\bigg\}. \label{smallact}
\end{align}
The Hamiltonian $H_m$, defined in (\ref{abbreviation}), can be easily diagonalized through a unitary Bogoliubov transformation
\begin{equation}
H_m=U_m\left(\begin{array}{cc}\epsilon_m & 0 \\0 & -\epsilon_m \end{array} \right)U_m^\dagger,\quad{}U_m=\left(\begin{array}{cc}\text{cos}\theta_m & i\text{sin}\theta_m \\i\text{sin}\theta_m & \text{cos}\theta_m \end{array} \right). \label{Bogo}
\end{equation}
In this expression
\begin{equation}
\epsilon_m=2\sqrt{\left(h-\text{cos}\frac{2\pi}{N}\left(m+\frac{1}{2}\right)\right)^2+\left(r\text{sin}\frac{2\pi}{N}\left(m+\frac{1}{2}\right)\right)^2}
\end{equation}
and 
\begin{equation}
\text{tan}\left(2\theta_m\right)=\frac{r\text{sin}\frac{2\pi}{N}\left(m+\frac{1}{2}\right)}{h-\text{cos}\frac{2\pi}{N}\left(m+\frac{1}{2}\right)}.
\end{equation}
By making the change of variables 
\begin{equation}
\left(\begin{array}{cc}c_m \\\bar{c}_{N-m-1}\end{array} \right)=U_m\left(\begin{array}{cc}\xi_m \\\bar{\xi}_{N-m-1}\end{array} \right) \label{changevar}
\end{equation}
the action (\ref{smallact}) can be written in the following form
\begin{equation}
S_m[J]=\int_{-\infty}^{\infty}d\tau\left(\bar{\eta}_m{D}_m\eta_m-i\bar{\mu}_m\eta_m-i\bar{\eta}_m\mu_m\right), \label{actionm}
\end{equation}
where 
\begin{equation}
\eta_m=\left(\begin{array}{cc}\xi_m \\\bar{\xi}_{N-m-1}\end{array}. \right),\quad{D}_m=\left(\begin{array}{cc}\partial_\tau+\epsilon_m & 0 \\0 & \partial_\tau-\epsilon_m \end{array} \right) \label{morechangevar}
\end{equation}
and
\begin{equation}
\bar{\mu}_m=\left(\begin{array}{cc}\bar{\lambda}_m & -\lambda_{N-m-1} \end{array} \right){U_m}.
\end{equation}
Before proceeding, it is worth noting that the change of variables (\ref{changevar}) and the subsequent diagonalization is permitted by the symmetric form of the discrete time lattice structure which defines the path integral. On the contrary, if the asymmetric discrete form had been kept, this change would not be possible. Written in this form, the integrals in (\ref{observation}) can be easily calculated. In the limit $\beta\rightarrow\infty$ which we are interested in, the result reads 
\begin{equation}
Z_m[J]=Z_m[0]\text{exp}\left\{-\int_{-\infty}^{\infty}d\tau\int_{-\infty}^{\infty}d\tau'\bar{\mu}_m(\tau)G_m(\tau-\tau')\mu_m(\tau')\right\}, \label{genfunpropag}
\end{equation}
with 
\begin{equation*}
G_m\equiv{}D_m^{-1}=\left(\begin{array}{cc}G^{(+)}_m & 0 \\0 & G^{(-)}_m \end{array} \right)
\end{equation*}
and
\begin{equation}
G_m^{(+)}(\tau-\tau')=\Theta(\tau-\tau')e^{-(\tau-\tau')\epsilon_m},\quad{}G_m^{(-)}(\tau-\tau')=-\Theta(\tau'-\tau)e^{-(\tau'-\tau)\epsilon_m}. \label{propag}
\end{equation}
The Green's function $G_m^{(+)}$, which propagates the $m$ modes, has been chosen to obey casuality: $G^{(+)}(\tau-\tau')=0$ for $\tau-\tau'<0$. It is, in fact, the antiperiodic function $G_m^{(+)}(\tau-\tau')=\left[\Theta(\tau-\tau')-\left(1+e^{\beta\epsilon_m}\right)^{-1}\right]e^{-(\tau-\tau')\epsilon_m}$ at the limit $\beta\rightarrow\infty$. The advanced function $G_m^{(-)}(\tau-\tau')$ propagates the $N-m-1$ conjugate modes backwards, and obeys the boundary condition $G_m^{(-)}(\tau-\tau')=0$ for $\tau-\tau'>0$. As expected, it is the $\beta\rightarrow\infty$ limit of the antiperiodic Green's function \\$G_m^{(-)}(\tau-\tau')=\left[\left(1+e^{\beta\epsilon_m}\right)^{-1}-\Theta(\tau'-\tau)\right]e^{-(\tau'-\tau)\epsilon_m}$. Note that, according to our prescription, the $\Theta$ function appearing in (\ref{propag}) is the Heaviside step function, for which $\Theta(0)=1/2$.
In Eq. (\ref{genfunpropag}), the subsystem's partition function $Z_m[0]$ appears as a normalization factor. The calculation of this factor and, consequently, the calculation of the partition function of the whole system can be easily performed \cite{1} yielding the result
\begin{equation}
Z(\beta)=\prod_{m=0}^{N-1}2\text{cosh}(\beta\epsilon_m/2).
\end{equation}
By acting with the functional derivatives on the generating functional (\ref{genfunpropag}), it is an easy task to compute the following expressions, that are the basis for all correlation functions:
\begin{align}
\begin{split}
&\braket{\hat{T}\left(\hat{\psi}_b(\tau_2)\hat{\psi}^\dagger_a(\tau_1)\right)}=\\
&=\frac{1}{N}\sum_{m=0}^{N-1}e^{\frac{2\pi{i}}{N}\left(m+\frac{1}{2}\right)(b-a)}\left(\text{cos}^2\theta_mG^{(+)}_m(\tau_2-\tau_1)+\text{sin}^2\theta_mG^{(-)}_m(\tau_2-\tau_1)\right) \label{cor1}
\end{split}
\end{align}
\begin{align}
\begin{split}
&\braket{\hat{T}\left(\hat{\psi}^\dagger_b(\tau_2){\hat{\psi}}_a(\tau_1)\right)}=\\
&=-\frac{1}{N}\sum_{m=0}^{N-1}e^{\frac{2\pi{i}}{N}\left(m+\frac{1}{2}\right)(b-a)}\left(\text{cos}^2\theta_mG^{(+)}_m(\tau_1-\tau_2)+\text{sin}^2\theta_mG^{(-)}_m(\tau_1-\tau_2)\right) \label{cor2}
\end{split}
\end{align}
\begin{align}
\begin{split}
&\braket{\hat{T}\left(\hat{\psi}^\dagger_b(\tau_2)\hat{{\psi}}^\dagger_a(\tau_1)\right)}=\\
&=\frac{i}{2N}\sum_{m=0}^{N-1}e^{\frac{2\pi{i}}{N}\left(m+\frac{1}{2}\right)(b-a)}\text{sin}(2\theta_m)\left(G^{(+)}_m(\tau_1-\tau_2)-G^{(-)}_m(\tau_1-\tau_2)\right) \label{cor3}
\end{split}
\end{align}
\begin{align}
\begin{split}
&\braket{\hat{T}\left(\hat{\psi}_b(\tau_2){\hat{\psi}}_a(\tau_1)\right)}=\\
&=-\frac{i}{2N}\sum_{m=0}^{N-1}e^{\frac{2\pi{i}}{N}\left(m+\frac{1}{2}\right)(b-a)}\text{sin}(2\theta_m)\left(G^{(+)}_m(\tau_2-\tau_1)-G^{(-)}_m(\tau_2-\tau_1)\right). \label{cor4}
\end{split}
\end{align}
Note that by interchanging $\tau_1,\tau_2$ and $a,b$ in (\ref{cor2}) and comparing the result with (\ref{cor1}) we deduce that $\braket{\hat{T}\left(\hat{\psi}_b(\tau_2)\hat{\psi}^\dagger_a(\tau_1)\right)}=-\braket{\hat{T}\left(\hat{\psi}^\dagger_a(\tau_1)\hat{\psi}_b(\tau_2)\right)}$, which is anticipated since we are computing the correlation functions between fermionic operators. Furthermore, it is easy to see that 
$\braket{\hat{T}\left(\hat{\psi}_b(\tau_2){\hat{\psi}}_a(\tau_1)\right)}^*=\braket{\hat{T}^\dagger\left(\hat{\psi}^\dagger_a(\tau_1)\hat{\psi}^\dagger_b(\tau_2)\right)}$. In the equal time case, due to Eq. (\ref{eq}), correlation functions (\ref{cor1}) and (\ref{cor2}) receive an extra  $+\frac{1}{2}\delta_{ab}$ contribution. 
\newline
\newline
Due to the quadratic nature of the Hamiltonian in hand, these correlation functions contain all the information needed for the analysis of the system. As a first example we calculate the transverse magnetization $\braket{\sigma^z}$ which is a site-independent quantity due to the translational invariance of the system
\begin{equation}
\braket{\sigma^z}=2\frac{\delta^2\text{ln}Z[J]}{\delta\bar{J}_j(\tau)\delta{J}_j(\tau)}\bigg|_{J=0}=\frac{1}{N}\sum_{m=0}^{N-1}\text{cos}(2\theta_m).
\end{equation} 
At the thermodynamic limit $N\rightarrow\infty$ we get the known \cite{28.9} result
\begin{equation}
\braket{\sigma^z}=\frac{1}{\pi}\int_{0}^{\pi}d\phi\frac{|h-\text{cos}\phi|}{\sqrt{(h-\text{cos}\phi)^2+(r\text{sin}\phi)^2}},
\end{equation} 
which confirms that the correct functional operations representing the spin operators are those given in Eq. (\ref{operat}).
From the physical point of view, more interesting is the connected time-dependent correlation function
\begin{equation}
\braket{\hat{T}\left(\sigma_j^z(\tau_2)\sigma_k^z(\tau_1)\right)}_c=4\frac{\delta^4\text{ln}Z[J]}{\delta\bar{J}_j(\tau_2)\delta{J}_j(\tau_2)\delta\bar{J}_k(\tau_1)\delta{J}_k(\tau_1)}\bigg|_{J=0}. \label{spinspincorel}
\end{equation}
To find the real time result, we perform the Wick rotation $\tau\rightarrow{it}.$ The calculation of the correlator (\ref{spinspincorel}) is then quite simple and yields the following result:
\begin{equation}
\braket{\hat{T}\left(\sigma_j^z(t_2)\sigma_k^z(t_1)\right)}_c=A_{jk}(|t_2-t_1|)+B_{jk}(|t_2-t_1|), \label{spinrestim}
\end{equation}
where in the last expression we used the abbreviations
\begin{align}
\begin{split}
&A_{jk}(|t|)=\left\{\frac{2}{N}\sum_{m=0}^{N-1}e^{-i|t|\epsilon_m}\text{cos}^2\theta_m\text{cos}\left(\phi_m{l}\right)\right\}\left\{\frac{2}{N}\sum_{m=0}^{N-1}e^{-i|t|\epsilon_m}\text{sin}^2\theta_m\text{cos}\left(\phi_m{l}\right)\right\},\\
&B_{jk}(|t|)=\left\{\frac{1}{N}\sum_{m=0}^{N-1}e^{-i|t|\epsilon_m}\text{sin}2\theta_m\text{sin}\left(\phi_m{l}\right)\right\}^2,\\
&\phi_m=\frac{2\pi}{N}\left(m+\frac{1}{2}\right),\quad{}l=j-k. \label{funct}
\end{split}
\end{align}
By taking into account that $G_m^{(+)}(0)=-G_m^{(-)}(0)=1/2$ and setting, for reasons of comparison $r=1$, we get the static result 
\begin{equation}
\braket{\sigma_j^z\sigma_k^z}_c=-\Sigma(l)\Sigma(-l),\quad\Sigma(l)=hR(l)-R(l+1), \label{spinres}
\end{equation}
with 
\begin{equation}
R(l)=\frac{1}{N}\sum_{m=0}^{N-1}\frac{\text{cos}(\phi_ml)}{\sqrt{(h-\text{cos}\phi_m)^2+\text{sin}^2\phi_m}}\underset{N\rightarrow\infty}{=}\frac{1}{\pi}\int_{0}^{\pi}d\phi\frac{\text{cos}(\phi{l})}{\sqrt{(h-\text{cos}\phi)^2+\text{sin}^2\phi}}.
\end{equation}
The reason we derived the result (\ref{spinres}), a result that is known for a long time \cite{28.9}\footnote{The sign difference in $h$ is due to the sign difference of the magnetic field used in the Hamiltonian of the cited paper.}, is that it is a strong indication that the correct way to define path integration over fermionic coherent states is through the Faddeev-Jackiw method. 
\newline
\newline
At the thermodynamic limit $N\rightarrow\infty$ we get for the the functions (\ref{funct}), which form the correlator (\ref{spinrestim}),
\begin{align}
\begin{split}
&A_{jk}(|t|)=\frac{1}{\pi^2}\int_{0}^{\pi}d\phi\int_{0}^{\pi}d\phi'\frac{e^{-i|t|(\epsilon(\phi)+\epsilon(\phi'))}}{\epsilon(\phi)\epsilon(\phi')}\times\\
&\times\left[\epsilon(\phi)+2(h-\text{cos}\phi)\right]\left[\epsilon(\phi')-2(h-\text{cos}\phi')\right]\text{cos}(\phi{l})\text{cos}{(\phi'l)},\\
&B_{jk}(|t|)=\frac{4}{\pi^2}\int_{0}^{\pi}d\phi\int_{0}^{\pi}d\phi'\frac{e^{-i|t|(\epsilon(\phi)+\epsilon(\phi'))}}{\epsilon(\phi)\epsilon(\phi')}\text{sin}\phi\text{sin}{\phi'}\text{sin}(\phi{l})\text{sin}(\phi'{l}).
\end{split}
\end{align}
Here, $\epsilon(\phi)$ is the continuum limit of the discrete energy $\epsilon_m$
\begin{equation}
\epsilon(\phi)=2\sqrt{(h-\text{cos}\phi)^2+(r\text{sin}\phi)^2}.
\end{equation}
As it is obvious, the correlator (\ref{spinrestim}), and consequently entanglement, disappears at the limit $t\rightarrow\infty$ due to strong oscillations, except from the critical point where $|h-\text{cos}\phi|=\lambda<<1$ and $\phi\sim\lambda$. Thus, the time after which correlations are strongly diminished scales as $t\sim{1/\lambda}$.
\newline
\newline
We can also compute the real time correlator of the Majorana operators (\ref{majora})
\begin{equation}
iB_{ba}(t_2,t_1)\equiv\braket{\hat{T}\left(\hat{\gamma}_{2b}(t_2)\hat{\gamma}_{2a-1}(t_1)\right)}. \label{majocor}
\end{equation}
Inserting the Wick rotated correlators (\ref{cor1})-(\ref{cor4}) into Eq. (\ref{majocor}) we get the real time correlator 
\begin{equation}
B_{ba}(t_2,t_1)=-\frac{1}{N}\sum_{m=0}^{N-1}e^{\frac{2\pi{i}}{N}\left(m+\frac{1}{2}\right)(b-a)-2i\theta_m}e^{-i|t_2-t_1|\epsilon_m}, \label{majocorres}
\end{equation}
which at the thermodynamic limit $N\rightarrow\infty$ becomes
\begin{equation}
B_l(t)\equiv{}B_{ba}(t_2,t_1)=-\frac{1}{2\pi}\int_{0}^{2\pi}d\phi{}e^{il\phi-2i\theta(\phi)}e^{-i|t|\epsilon(\phi)}, \quad{l}=b-a,\quad{t=t_2-t_1}. \label{majocorrestherm}
\end{equation}
In this expression the function $\theta(\phi)$ is defined as
\begin{equation}
2\theta(\phi)=\left\{\begin{array}{rl}\text{arctan}\frac{r\text{sin}\phi}{h-\text{cos}\phi} & ,\quad h-\text{cos}\phi>0,\\\text{arctan}\frac{r\text{sin}\phi}{h-\text{cos}\phi}+\pi &,\quad h-\text{cos}\phi<0.\end{array} \right. \label{cases}
\end{equation} 
For the $XX$ model $(r\rightarrow0)$ and $|h|\leq1$ the integral (\ref{majocorrestherm}) simplifies to
\begin{align}
B_{l}(t)&=\frac{1}{2\pi}\int_{0}^{2\pi}d\phi{}e^{il\phi}\text{sign}(\text{cos}\phi-h)e^{-2i|t||\text{cos}\phi-h|}=\nonumber\\
&=\frac{1}{\pi}\int_{-\phi_h}^{\phi_h}d\phi{}e^{il\phi}\text{cos}\left[2t(h-\text{cos}\phi)\right],\label{resrse}
\end{align}
where
$\phi_h$ is defined through the equation $cos\phi_h=h$.\newline
The evaluation of this integral yields the result
\begin{equation}
B_{l}(t)=\frac{2}{\pi}\sum_{k=-\infty}^{\infty}J_k(2t)\frac{\text{sin}\left[(l+k)\phi_h\right]}{l+k}\text{cos}\left(2ht-k\frac{\pi}{2}\right) \label{besselcor}
\end{equation}
with $J_k$ being the Bessel functions. For the static case, one does not have to repeat all computations again by taking into account Eq. (\ref{eq}), since when identifying the weighted quantity corresponding to $\hat{T}\left(\hat{\gamma}_{2b}(t)\hat{\gamma}_{2a-1}(t)\right)$ through the Faddeev-Jackiw prescription, all extra $1/2$ contributions cancel, giving the function ${\gamma}_{2b}(t){\gamma}_{2a-1}(t)$. Thus, we can just take the limit $t_1=t_2$ of Eq. (\ref{besselcor}) in which case only the $k=0$ term survives, giving
\begin{equation}
B_{ba}^{(eq)}(t)=\frac{2}{\pi}\frac{\text{sin}(l\phi_h)}{l}.
\end{equation}
Here we have defined the equal time correlation function as $B_{ba}^{(eq)}(t)\equiv{B}_{ba}(t,t)$.
At the vicinity of the critical value $h=1$ we find that $\phi_h\simeq\sqrt{2\lambda}$ where $\lambda\equiv1-h\rightarrow+0$. This value sets the scale after which strong oscillations diminish correlations. Thus, the correlation length behaves as $\xi\sim\lambda^{-\frac{1}{2}}\sim\lambda^{-\nu}$ giving the corresponding critical exponent $\nu=1/2$ \cite{21}. As $|t|\rightarrow\infty$, the correlator (\ref{besselcor}) goes to zero as $B_l\sim|t|^{-1/2}$, except for the critical vicinity where it oscillates as 
\begin{equation}
B_{l}(t)\simeq{}\frac{2}{\pi{l}}\text{sin}(\phi_hl)\text{cos}(2t\lambda).
\end{equation}
The time scale after which correlations are turned off due to the strong oscillations can easily be seen from the last equation being: $t\sim\lambda^{-1}\sim\xi^2$. Consequently, the dynamical critical exponent defined through equation $t\sim\xi^z$ is $z=2$ \cite{22}. Therefore, we expect  \cite{11,12,13} the entanglement entropy to scale as $S\sim\frac{1}{3}\text{log}\xi\sim\frac{1}{6}\text{log}t$ at the critical vicinity.

\subsection{Driven correlations}

In this section we examine the case of a time dependent transverse field $h=h(\tau)$ driving the evolution of an $XY$ spin chain. For the quantum Ising model ($r=1$) and for a field linearly dependent on time, the entanglement dynamics have been extensively studied \cite{14,15,30,31,32}, mainly by numerical methods. The present work is a contribution to the analytical methods available for the study of spin systems out of equilibrium. In the path integral representation, the calculation of the equal time vacuum state correlators of the $XY$ model can proceed irrespectively of the form the function $h(\tau)$ has.  Leaving the details of the calculation for Appendix B, it suffices here to note that in the presence of a time dependent magnetic field, both the eigenvalues $\epsilon_m$ and the matrix $U_m$, which diagonalizes the Hamiltonian in Eq. (\ref{abbreviation}), become time dependent quantities. As a consequence, the extra off-diagonal contribution
\begin{equation}
U^\dagger_m\partial_tU_m=i\dot{\theta}_m\sigma^x,\quad\dot{\theta}_m=-\dot{h}\frac{2r}{\epsilon_m^2}\text{sin}\left[\frac{2\pi}{N}\left(m+\frac{1}{2}\right)\right]=-\dot{h}\frac{2r}{\epsilon_m^2}\text{sin}\phi_m\quad(r\neq0)
\end{equation}
appears in the action (\ref{actionm}), a fact that makes the problem more involved. 
\newline
\newline
For the case of the $XX$ model ($r=0$) this is not the case since even though the angle $2\theta_m$ is not constant as it jumps between $0$ and $\pi$ according to Eq. (\ref{cases}), its time derivative is zero. Consequently we can repeat the steps we followed for the time-independent system without using any approximations. This way we get the exact result 
\begin{equation}
B_{ba}^{(eq)}(\tau)=\frac{1}{\pi}\int_{-\phi_h(\tau)}^{\phi_h(\tau)}d\phi{e}^{il\phi}=\frac{2}{\pi}\frac{\text{sin}\left(l\phi_h(\tau)\right)}{l},\quad{l=b-a}.
\end{equation}
\newline
For $r\neq0$ however, the calculation can also be carried out due to the quadratic nature of the Hamiltonian appearing in Eq. (\ref{xyham}). In Appendix B, the partition function of the driven system is found to have the form 
\begin{equation}
Z(\beta)=\prod_{m=0}^{N/2-1}\text{Det}(\tilde{D}_m),\label{partpart}
\end{equation}
where 
\begin{equation}
\tilde{D}_m\equiv\left(\begin{array}{cc} \partial_\tau+{\epsilon}_m(\tau) & i\dot{\theta}_m(\tau) \\i\dot{\theta}_m(\tau) & \partial_\tau-{\epsilon}_m(\tau) \end{array} \right)=D_m+i\dot{\theta}_m\sigma^x.\label{kinem}
\end{equation}
In the same Appendix we also prove that both the determinant of (\ref{kinem}) and the Green's function $\tilde{G}_m=\tilde{D}_m^{-1}$ can be calculated as a convergent series in powers of $\dot{\theta}_m$. As a result, the partition function (\ref{partpart}) can be recasted to the form
\begin{equation}
Z(\beta)=Z_0(\beta)e^{-\frac{1}{2}\sum_{m=0}^{N-1}E_m},\label{ppppaart}
\end{equation}
where 
\begin{equation}
Z_0(\beta)=\prod_{m=0}^{N-1}2\text{cosh}\left(\int_{-\beta/2}^{\beta/2}d\tau\epsilon_m(\tau)\right).\label{partt}
\end{equation}
The exponential factor in Eq. (\ref{ppppaart}) can then be expressed as a convergent series in $\dot{\theta}_m$
\begin{equation}
E_m=\frac{1}{2}\text{Tr}K_m^2+\frac{1}{4}\text{Tr}K_m^4+\dots
\end{equation}
where\footnote{In index notation
	\begin{equation*}
	\left({K}_m(\tau_2-\tau_1)\right)^a{}_b=i\left(G_m(\tau_2-\tau_1)\right)^a{}_c\left(\sigma^x\right)^{c}{}_b\dot{\theta}_m(\tau_1)
	\end{equation*}} ${K}_m=iG_m\dot{\theta}_m\sigma^x$. 
In this expression the trace symbol denotes tracing over time and matrix indices:
\begin{equation}
\text{Tr}\left(\dots\right)=\text{tr}\int_{-\beta/2}^{\beta/2}d\tau\braket{\tau|(\dots)|\tau}=\int_{-\beta/2}^{\beta/2}d\tau\sum_{aa}\braket{\tau|(\dots)_{aa}|\tau}.
\end{equation}
In the adiabatic regime, this function can be formally expanded with respect to the frequency parameter $\omega$ of the magnetic field's evolution, giving at the limit $\beta\rightarrow\infty$
\begin{align}
\begin{split}
E_m\simeq\frac{\omega}{2}\sum_{m=0}^{N-1}\int_{-\infty}^{\infty}{d\sigma}\frac{\dot{\theta}^2_m(\sigma)}{\epsilon_m(\sigma)}+\mathcal{O}(\omega^2).
\end{split}
\end{align}
The parameter $\omega$ measures the rate of change of the magnetic field and appears inside functions multiplied to $\tau$, making $\omega\tau$ dimensionless. In this result the change of variables $\sigma=\omega\tau$ has been made, such that an adiabatic expansion was possible.
To be concrete, we can consider the per site free energy 
\begin{equation}
-\frac{1}{N}\text{ln}Z(\beta)=-\frac{1}{N}\text{ln}Z_0(\beta)+\frac{1}{2N}\sum_{m=0}^{N-1}E_m.
\end{equation}
At the  thermodynamic limit $N\rightarrow\infty$, the term 
\begin{equation}
\frac{1}{N}\sum_{m=0}^{N-1}E_m\underset{N\rightarrow\infty}{\simeq}\frac{1}{2\pi}\int_{0}^{2\pi}d\phi{E}(\phi)
\end{equation}
is controlled by the poles that correspond to the zeros of the energy and turns out to be a temperature-independent constant both at the adiabatic and the sudden limit. This result supports the conclusion that, at these limits, the critical behaviour of the system is determined mainly by the partition function $Z_0(\beta)$.
In the same context, in Appendix B, we prove that the equal time correlator (\ref{majocor}) can be calculated through the relation 
\begin{equation}
B_{ba}^{(eq)}(\tau)\equiv-i\braket{\hat\gamma_{2b}(\tau)\hat{\gamma}_{2a-1}(\tau)}=-\frac{1}{N}\sum_{m=0}^{N-1}e^{\frac{2\pi{i}}{N}\left(m+\frac{1}{2}\right)l-2i\theta_m(\tau)}\text{Tr}(\Sigma\tilde{G}_m),\label{coooor}
\end{equation}
where 
\begin{equation}
\Sigma\equiv\sigma^z-i\sigma^y=\left(\begin{array}{cc} 1 & -1 \\1 & -1 \end{array} \right)
\end{equation}
and $l=b-a$.
Using this we find 
\begin{equation}
B^{(eq)}_{ba}(\tau)=-\frac{1}{N}\sum_{m=0}^{N-1}e^{\frac{2\pi{i}}{N}\left(m+\frac{1}{2}\right)l-2i\theta_m(\tau)}\left(1+\sum_{\nu=1}^{\infty}c^{(\nu)}_m(\tau)\right)\equiv{B_l^S}(\tau)+B_l^Q(\tau),\label{comp}
\end{equation}
the first term of which in the thermodynamic limit reads:
\begin{equation}
B_l^S(\tau)\underset{N\rightarrow\infty}{=}-\frac{1}{2\pi}\int_{0}^{2\pi}d\phi{e}^{i\phi{l}-2i\theta(\phi;\tau)}
\end{equation}
reproducing for each instantaneous value $h(\tau)$ the corresponding static results \cite{12}.
For the second term of Eq. (\ref{comp}), in the adiabatic regime $\omega\rightarrow0$, one can again calculate the zero temperature result ($\beta\rightarrow\infty$), which for a general magnetic field is
\begin{equation}
B_l^Q(\tau)\simeq{i}\frac{\omega^2}{4\pi}\int_{0}^{2\pi}d\phi\frac{{e}^{i\phi{l}-2i\theta(\phi;\sigma)}}{\epsilon(\phi;\sigma)}\partial_\sigma\left(\frac{\dot{\theta}_m(\phi;\sigma)}{\epsilon(\phi;\sigma)}\right)+\mathcal{O}(\omega^3)\label{bq}
\end{equation}
with $\sigma=\omega\tau$.
If we consider $h(\tau)=\omega\tau$ we can study this term more extensively, to understand the adiabatic and sudden limits.
\newline
\newline
In Appendix B we show that when the driving is slow enough and the magnetic field away from its critical value, $B_l^Q\underset{\omega\rightarrow0}{\sim}\mathcal{O}(\omega^2)$ is negligible in comparison to $B_l^S$ in Eq. (\ref{comp}). Thus, if the evolution is adiabatic, the "static" term $B_l^S$ controls the correlator. However, as the system approaches the critical region $|h-1|\sim|\phi|\underset{\omega\rightarrow0}{\sim}\mathcal{O}(\omega^{1/2})$, $B_l^Q$ becomes increasingly important and approaches its sudden limit. At the critical point $h=1$ it is very simple to find that
\begin{equation}
B_l^Q(1)\simeq\frac{i}{4}\int_{-\phi_0}^{\phi_0}d\phi{e}^{i\phi{l}-2i\theta(\phi;1)}=\frac{\text{sin}\left[\phi_0(2l+1)/2\right]}{2l+1},\quad\phi_0=\mathcal{O}(\omega^{1/2}).
\end{equation}
This behaviour sets the length $\xi\sim\omega^{-1/2}$ as the scale which characterizes the passing of the system through the critical point. This is in accordance with the so-called Kibble-Zurek mechanism (KZM) or the adiabatic-impulse-adiabatic approximation \cite{22,23}, which is based on the fact that the evolution of a system driven through a second order phase transition cannot be adiabatic near the critical point, irrespective of how slow the driving is. In KZM, time evolution is considered initially as adiabatic, becoming non-adiabatic near the critical point where the energy gap changes with a rate comparable to the energy gap itself: $|\dot{h}|/|h-1|\sim|h-1|\sim\mathcal{O}(\omega^{1/2})$. In such a case, the entanglement entropy is expected \cite{15} to behave as $S\simeq\frac{1}{12}\log_2\frac{1}{\omega}.$
 
 \section{Conclusions-Perspectives}
 In the present paper we have used path integration over fermionic coherent states to analyse quantum correlations in a paradigmatic spin model. We discussed the construction of the relevant path integral and we adopted the Faddeev-Jackiw method to avoid possible pitfalls. We calculated the time-dependent vacuum expectation values needed for the calculation of the entanglement entropy and we confirmed that the correct static limit is reproduced. In the last section we examined the case of a general time-dependent magnetic field, and in the case that the system is driven through the critical point we confirmed that our results are consistent with the Kibble-Zurek mechanism. The aim of the current work is not only to present new analytical results regarding the dynamics of the $XY$ model, but also to present a novel way to analyse closed spin-chain systems in general. In a forthcoming work we shall use the path integral method to derive the reduced dynamics of open spin-chain systems.
 
 \section*{Acknowledgements}
 This research is co-financed by Greece and the European Union (European Social Fund- ESF) through the Operational Programme <<Human Resources Development, Education and Lifelong Learning>> in the context of the project ''Strengthening Human Resources Research Potential via Doctorate Research'' (MIS-5000432), implemented by the State Scholarships Foundation (IKY)
 
 \appendix

\section{Functional integration of $\hat{H}=-\omega\hat{\vec{S}}_1\cdot\hat{\vec{S}}_2$}

In this Appendix, we examine a simple system, the dynamics of which can be described by the Hamiltonian $\hat{H}=-\omega\hat{\vec{S}}_1\cdot\hat{\vec{S}}_2$. The partition function of this system is known to be
\begin{equation}
Z=\text{Tr}\left[e^{-\beta\hat{H}}\right]=e^{-3\beta\omega/4}+3e^{\beta\omega/4}.
\end{equation}
Using the Jordan-Wigner transformation, the Hamiltonian in hand assumes the form
\begin{equation}
\hat{H}=-\frac{\omega}{4}\left[\left(\hat{\psi}^\dagger_1-\hat{\psi}_1\right)\left(\hat{\psi}^\dagger_2+\hat{\psi}_2\right)+\left(\hat{\psi}^\dagger_2-\hat{\psi}_2\right)\left(\hat{\psi}^\dagger_1+\hat{\psi}_1\right)+\left(1-2\hat{\psi}^\dagger_1\hat{\psi}_1\right)\left(1-2\hat{\psi}^\dagger_2\hat{\psi}_2\right)\right],
\end{equation}
which in terms of Majorana operators reads as
\begin{equation}
\hat{H}=i\frac{\omega}{4}\left(\hat{\gamma}_2\hat{\gamma}_3+\hat{\gamma}_4\hat{\gamma}_1-i\hat{\gamma}_1\hat{\gamma}_2\hat{\gamma}_3\hat{\gamma}_4\right). \label{app1hammaj}
\end{equation}
According to the Faddeev-Jackiw quantization scheme, the classical function entering the Majorana path integral representation is the classical counterpart of (\ref{app1hammaj})
\begin{equation}
H_M=i\frac{\omega}{4}\left({\gamma}_2{\gamma}_3+{\gamma}_4{\gamma}_1-i{\gamma}_1{\gamma}_2{\gamma}_3{\gamma}_4\right), \label{app1hammajvar}
\end{equation}
which in turn translates to the complex Grassmann variables as
\begin{equation}
H=-\frac{\omega}{2}\left(\bar{\zeta}_1\zeta_2+\bar{\zeta}_2\zeta_1+2|\zeta_1|^2|\zeta_2|^2\right). \label{app1hammajcom}
\end{equation}
Thus, the integral to be evaluated is
\begin{equation}
Z=\left(\prod_{j=1}^{2}\int_{AP}\mathcal{D}\bar{\zeta}_j\mathcal{D}\zeta_je^{-\int_{-\frac{\beta}{2}}^{\frac{\beta}{2}}d\tau\bar{\zeta}_j\dot{\zeta}_j}\right)\text{exp}\left[\frac{\omega}{2}\int_{-\frac{\beta}{2}}^{\frac{\beta}{2}}d\tau\left(\bar{\zeta}_1\zeta_2+\bar{\zeta}_2\zeta_1+2|\zeta_1|^2|\zeta_2|^2\right)\right] \label{app1intgen}.
\end{equation}
It is convenient to perform a change of variables, induced through the unitary transformation:
\begin{equation}
\left(\begin{array}{cc}\zeta_1 \\\zeta_{2}\end{array} \right)=\frac{1}{\sqrt{2}}\left(\begin{array}{cc}1 & 1\\1 & -1\end{array} \right)\left(\begin{array}{cc}\eta_1 \\\eta_{2}\end{array} \right).
\end{equation}
After this change, the integral (\ref{app1intgen}) is recasted to the form
\begin{equation}
Z=\left(\prod_{j=1}^{2}\int_{AP}\mathcal{D}\bar{\eta}_j\mathcal{D}\eta_je^{-\int_{-\frac{\beta}{2}}^{\frac{\beta}{2}}d\tau\bar{\eta}_j\dot{\eta}_j}\right)exp\left[\frac{\omega}{2}\int_{-\frac{\beta}{2}}^{\frac{\beta}{2}}d\tau\left(|{\eta}_1|^2-|{\eta}_2|^2+2|\eta_1|^2|\eta_2|^2\right)\right] \label{app1intgen2}.
\end{equation}
We proceed by performing the integration over the first Grassmann field 
\begin{align}
Z_1&=\int_{AP}\mathcal{D}\bar{\eta}_1\mathcal{D}\eta_1\text{exp}\left[-\int_{-\frac{\beta}{2}}^{\frac{\beta}{2}}d\tau\bar{\eta}_1\left(\partial_\tau-\frac{\omega}{2}-\omega|\eta_2|^2\right)\eta_1\right]=\nonumber\\
&=2\text{cosh}\left(\frac{\beta\omega}{4}+\frac{\omega}{2}\int_{-\frac{\beta}{2}}^{\frac{\beta}{2}}d\tau|\eta_2|^2\right). \label{result1}
\end{align}
To arrive to the last result, we have relied on the symmetric prescription for the underlying lattice structure.
Inserting Eq. (\ref{result1}) into Eq. (\ref{app1intgen2}), we get the correct quantum result:
\begin{align}
Z&=e^{{\beta\omega}/{4}}\int_{AP}\mathcal{D}\bar{\eta}_2\mathcal{D}\eta_2e^{-\int_{-\frac{\beta}{2}}^{\frac{\beta}{2}}d\tau\bar{\eta}_2\dot{\eta}_2}+e^{-{\beta\omega}/{4}}\int_{AP}\mathcal{D}\bar{\eta}_2\mathcal{D}\eta_2e^{-\int_{-\frac{\beta}{2}}^{\frac{\beta}{2}}d\tau\bar{\eta}_2\left(\partial_\tau+\omega\right)\eta_2}=\nonumber\\
&=e^{-3\beta\omega/4}+3e^{\beta\omega/4}.
\end{align}

\section{Calculation of equal time correlation function}
In this Appendix we prove the basic relations of Section 3.2.
\newline
\newline
Due to the time dependence of the magnetic field, the partition function of the system assumes the form
\begin{equation}
Z(\beta)=\prod_{m=0}^{N/2-1}Z_m(\beta),
\end{equation}
where 
\begin{equation}
Z_m(\beta)=\underset{AP}{\int}\mathcal{D}\bar{c}_m\mathcal{D}c_m\mathcal{D}\bar{c}_{N-1-m}\mathcal{D}c_{N-1-m}\text{exp}\left\{-\int_{-\beta/2}^{\beta/2}d\tau\bar{\xi}_m\left(D_m+i\dot{\theta}_m\sigma^x\right)\xi_m\right\}.
\end{equation}
Thus 
\begin{equation}
Z(\beta)=\prod_{m=0}^{N/2-1}\text{Det}\left(D_m+i\dot{\theta}_m\sigma^x\right)\equiv\prod_{m=0}^{N/2-1}\text{Det}\left(\tilde{D}_m\right).\label{partit}
\end{equation}
For the calculation of the functional determinant we can now write
\begin{equation}
\text{Det}\left(D_m+i\dot{\theta}_m\sigma^x\right)=\text{Det}\left(D_m\right)\text{Det}\left(\mathbb{1}+K_m\right),\label{detet}
\end{equation}
with ${K}_m=iG_m\dot{\theta}_m\sigma^x$
and
\begin{equation}
G_m={D}_m^{-1}=\left(\begin{array}{cc} G_m^{(+)} & 0 \\0 & G_m^{(-)} \end{array} \right).
\end{equation}
In the present case, $G_m^{(+)}$ is the antiperiodic retarded Green's function:\begin{equation}
G_m^{(+)}(\tau,\tau')=\left[\Theta(\tau-\tau')-\left(1+e^{\int_{-\beta/2}^{\beta/2}d\tau\epsilon(\tau)}\right)^{-1}\right]e^{-\int_{\tau'}^{\tau}d\tilde{\tau}\epsilon(\tilde{\tau})}
\end{equation}
and $G_m^{(-)}(\tau,\tau')=-G_m^{(+)}(\tau',\tau)$ is the antiperiodic advanced Green's function. Thus, the first factor in Eq. (\ref{detet})  is immediately found to be
\begin{equation}
\text{Det}\left(D_m\right)=e^{-\text{Trln}D_m^{-1}}=4\text{cosh}^2\int_{-\beta/2}^{\beta/2}d\tau\epsilon_m(\tau),
\end{equation}
and as a result, its contribution to the partition function (\ref{partit}) is
\begin{equation}
Z_0(\beta)=\prod_{m=0}^{N/2-1}\text{Det}\left(D_m\right)=\prod_{m=0}^{N-1}2\text{cosh}\int_{-\beta/2}^{\beta/2}d\tau\epsilon_m(\tau) 
\end{equation}
as noted in Eq. (\ref{partt}).
For the calculation of the functional determinant we rewrite
\begin{equation}
\text{Det}\left(\mathbb{1}+K_m\right)=\text{Det}\left(\mathbb{1}+\lambda{}K_m\right)\big|_{\lambda=1},
\end{equation}
so we can take advantage of the fact that $\text{Det}\left(\mathbb{1}+\lambda{}K_m\right)$ can be expanded as a convergent power series with respect to $\lambda$ \cite{32.5}, as
\begin{equation}
\text{Det}\left(\mathbb{1}+\lambda{}K_m\right)=\sum_{n=0}^{\infty}\frac{1}{n!}d^{(m)}_n\lambda^n,
\end{equation}
with 
\begin{equation}
d^{(m)}_n=\text{det}\left(\begin{array}{ccccc} \text{Tr}K_m & n-1 &  & \\\text{Tr}K^2_m & \text{Tr}K_m & n-2 &  & \\
. & . & . & . & \\
. & . & . & . & \\
\text{Tr}K^n_m & \text{Tr}K^{n-1}_m & . & . & \text{Tr}K_m\end{array} \right),\quad{n\geq1} \label{dn}
\end{equation}
and $d_0=1$.
This series is convergent $\forall$ $\lambda$ as long as $K_m$ is well behaved, i.e. when $||K_m||^2<\infty$ which is the case here since
\begin{equation}
||K_m||^2=\int_{-\beta/2}^{\beta/2}d\tau_1\int_{-\beta/2}^{\beta/2}d\tau_2\dot{\theta}^2_m(\tau_2)\left[|G^{(+)}(\tau_1-\tau_2)|^2+|G^{(-)}(\tau_1-\tau_2)|^2\right].\label{compli}
\end{equation} 
It can be easily confirmed that $\text{Tr}K_m^{2\nu-1}=0$, $\nu=1,2,\dots$. Thus, in the coefficients (\ref{dn}) only the even powers of $K_m$ contribute. Taking then the limit of $\lambda=1$ we find for the exponential factor appearing in Eq. (\ref{ppppaart}):
\begin{align}
\begin{split}
E_m&=-\text{logDet}\left(\mathbb{1}+K_m\right)=-\text{log}\left\{\sum_{\nu=0}^{\infty}\frac{1}{n!}d^{(m)}_n\right\}=\\
&=-\log\left\{1-\frac{1}{2}\text{Tr}K_m^2-\frac{1}{4}\left(\text{Tr}K_m^4-\frac{1}{2}\left(\text{Tr}K_m^2\right)^2\right)+\dots\right\}\\
&=\frac{1}{2}\text{Tr}K_m^2+\frac{1}{4}\text{Tr}K_m^4+\dots{}. \label{energy}
\end{split}
\end{align} 
This conclusion is also valid at the limit $\beta\rightarrow\infty$ and by analytic continuation can be extended to the real time formulation.
\newline
\newline
The contribution of this term to the partition function of the system can be investigated by writing 
\begin{equation}
-\frac{1}{N}\text{ln}Z(\beta)=-\frac{1}{N}\text{ln}Z_0(\beta)+\frac{1}{2N}\sum_{m=0}^{N-1}E_m.\label{beforeprobe}
\end{equation}
To probe the adiabatic and the sudden limits it is enough to consider the first term in the last factor of Eq. (\ref{beforeprobe}), and understand that $\tau$ appears inside $h(\tau)$ multiplied with a parameter $\omega$, such as $\omega\tau$ is a dimensionless quantity. Thus, $\omega$ is a parameter measuring the rate of change of the magnetic field, which gives the adiabatic and sudden limits at $\omega\rightarrow0$ and $\omega\rightarrow\infty$ respectively. After the rescaling $\omega{\tau}=\sigma$, and taking the limit $\beta\rightarrow\infty$, the aforementioned term reads:
\begin{equation}
\frac{1}{4N}\sum_{m=0}^{N-1}\text{Tr}K_m^2=\frac{1}{2N}\sum_{m=0}^{N-1}\int_{-\infty}^{\infty}d\sigma{}\dot{\theta}_m(\sigma)\int_{-\infty}^{\sigma}d\sigma'\dot{\theta}_m(\sigma')e^{-\frac{2}{\omega}\int_{\sigma'}^{\sigma}d\tilde{\sigma}\epsilon_m(\tilde{\sigma})},
\end{equation}
where the now $\omega$ independent, rescaled quantities are defined as
\begin{equation}
\dot{\theta}_m(\sigma)=\frac{1}{\omega}\dot{\theta}(\tau),\quad\epsilon_m(\sigma)=2\sqrt{(h(\sigma/\omega)-\text{cos}\phi_m)^2+(r\text{sin}\phi_m)^2}.\label{resc}
\end{equation}
At the adiabatic limit and as long as the system is not in the critical vicinity, we can use repeatedly the identity 
\begin{equation}
e^{-\frac{2}{\omega}\int_{\sigma'}^{\sigma}d\tilde{\sigma}\epsilon_m}=\frac{\omega}{2}\frac{1}{\epsilon_m(\sigma')}\partial_{\sigma'}e^{-\frac{2}{\omega}\int_{\sigma'}^{\sigma}d\tilde{\sigma}\epsilon_m},\label{id}
\end{equation}
to find 
\begin{align}
\begin{split}
\frac{1}{4N}\sum_{m=0}^{N-1}\text{Tr}K_m^2&\underset{\omega\rightarrow0,\beta\rightarrow\infty}{\simeq}\frac{\omega}{4N}\sum_{m=0}^{N-1}\int_{-\infty}^{\infty}{d\sigma}\frac{\dot{\theta}^2_m(\sigma)}{\epsilon_m(\sigma)}+\mathcal{O}(\omega^2).
\end{split}
\end{align}
While this result is a good approximation when the system is in the adiabatic regime and away from the critical point, it is interesting to study how it behaves for a system that passes from $h(\tau)=1$, as in the case of linear driving $h(\tau)=\omega\tau$, where at time $\sigma=1$ the critical point is found:
\begin{align}
\begin{split}
\frac{1}{4N}\sum_{m=0}^{N-1}\text{Tr}K_m^2&\underset{\omega\rightarrow0,\beta\rightarrow\infty}{\simeq}\frac{\omega{r^2}}{N}\sum_{m=0}^{N-1}\text{sin}^2\phi_m\int_{-\infty}^{\infty}\frac{d\sigma}{\epsilon_m^5(\sigma)}+\mathcal{O}(\omega^2)=\\
&=\frac{\omega}{24N}\sum_{m=0}^{N-1}\frac{1}{r^2\text{sin}^2\phi_m}+\mathcal{O}(\omega^2).
\end{split}
\end{align}
This result indicates that the main contribution to the integral \\ $\frac{1}{N}\sum_{m=0}^{N-1}E_m\underset{N\rightarrow\infty}{\simeq}\frac{1}{2\pi}\int_{0}^{2\pi}d\phi{E}(\phi)$ appearing in the exponential
of Eq. (\ref{ppppaart}), at the thermodynamic limit $N\rightarrow\infty$, is controlled by the poles produced by the zeros of the energy, forcing the successible $\sigma$ integrations to be restricted in a small region of width $\delta\sigma\sim\omega^{1/2}$ around the critical point $\sigma=1$ where $|\text{sin}\phi_m|=\mathcal{O}(\omega^{1/2})$, $\epsilon_m=\mathcal{O}(\omega^{1/2})$ and $\dot{\theta}_m=\mathcal{O}(\omega^{-1/2})$. In this regime $\int_{\sigma'}^{\sigma}d\tilde{\sigma}{\epsilon}_m(\tilde{\sigma})/\omega\simeq\mathcal{O}(1)$ and $\text{Tr}K_m^2\simeq\mathcal{O}(1)$. However, in such a case, all the terms in the sum (\ref{energy}) are almost constant, $\text{Tr}K_m^{2\nu}\simeq\mathcal{O}(1)$ and they all contribute to the sum, leading to a finite, temperature-independent result. At the sudden limit $\omega\rightarrow\infty$ the main contribution is again controlled by the poles of the integral in the $N\rightarrow\infty$ limit of (\ref{ppppaart}), yielding a temperature-independent result.
\newline
\newline
To prove Eq. (\ref{coooor}) of the main text we begin by writing the equal time correlaton function (\ref{majocor}) in the form 
\begin{equation}
B_{ba}^{(eq)}(\tau)=\braket{\hat{{\psi}}^\dagger_b\hat{\psi}_a-\hat{\psi}_b\hat{{\psi}}^\dagger_a+\hat{{\psi}}^\dagger_b\hat{{\psi}}^\dagger_a-\hat{\psi}_b\hat{\psi}_a}(\tau).
\end{equation}
All the operators in the last expression are defined at the same moment $\tau$. Following the standard procedure, we Fourier transform the fermionic operators (in analogy to the fermionic variables of Eq. (\ref{Fourier})) and get the following expression for the correlator
\begin{equation}
B_{ba}^{(eq)}(\tau)=\frac{1}{N}\sum_{m=0}^{N-1}e^{\frac{2\pi{i}}{N}\left(m+\frac{1}{2}\right)(b-a)}\braket{\left(\begin{array}{cc}\hat{\bar{c}}_m & \hat{c}_{N-m-1} \end{array} \right)\left(\begin{array}{cc} 1 & -1 \\1 & -1 \end{array} \right)\left(\begin{array}{cc}\hat{c}_m \\ 
	\hat{\bar{c}}_{N-m-1}\end{array} \right)}. 
\end{equation}
As noted in Section 3.1 the correct Hamiltonian symbol  at the level of path integration, when two same index operators appear as $\hat{\bar{c}}_m\hat{c}_m$, is not the $\mathbb{C}$-number quantity $\bar{c}_m{c}_m$, but instead $\bar{c}_m{c}_m+1/2$, which makes the correct path integral identification non-trivial. In this case thus, we should consider mapping the two non-diagonal terms with different indices trivially, but the two diagonal terms as
\begin{equation*}
\hat{\bar{c}}_m\hat{c}_m\rightarrow\bar{c}_mc_m+1/2,
\end{equation*}
\begin{equation*}
-\hat{c}_{N-m-1}\hat{\bar{c}}_{N-m-1}=\hat{\bar{c}}_{N-m-1}\hat{c}_{N-m-1}-1\rightarrow\bar{c}_{N-m-1}{c}_{N-m-1}-1/2.
\end{equation*}
It is evident that in this case the extra contributions cancel, but since this was not the case for the spin correlation functions computed in	Section 3.1 it must always be taken into account when considering equal time correlation functions. Thus, the quantity we want to calculate is
\begin{align}
\begin{split}
B_{ba}^{(eq)}(\tau)&=\frac{1}{N}\sum_{m=0}^{N-1}e^{\frac{2\pi{i}}{N}\left(m+\frac{1}{2}\right)(b-a)}\underset{AP}{\int}\mathcal{D}\bar{c}_m\mathcal{D}c_m\mathcal{D}\bar{c}_{N-1-m}\mathcal{D}c_{N-1-m}\\
&\left(\begin{array}{cc}{\bar{c}}_m & {c}_{N-m-1} \end{array} \right)\left(\begin{array}{cc} 1 & -1 \\1 & -1 \end{array} \right)\left(\begin{array}{cc}{c}_m \\ 
	{\bar{c}}_{N-m-1}\end{array} \right)\text{exp}\left\{-\int_{-\beta/2}^{\beta/2}d\tau\bar{\eta}_m\left(D_m+i\dot{\theta}_m\sigma^x\right)\eta_m\right\}.
\end{split}
\end{align}
By making the change of variables indicated in Eqs. (\ref{changevar}) and (\ref{morechangevar}), we easily confirm that:
\begin{align}
\begin{split}
B_{ba}^{(eq)}(\tau)&=\frac{1}{N}\sum_{m=0}^{N-1}e^{\frac{2\pi{i}}{N}\left(m+\frac{1}{2}\right)(b-a)-2i\theta_m}\underset{AP}{\int}\mathcal{D}\bar{c}_m\mathcal{D}c_m\mathcal{D}\bar{c}_{N-1-m}\mathcal{D}c_{N-1-m}\\
&\left(\bar{\eta}_m\Sigma\eta_m\right)\text{exp}\left\{-\int_{-\beta/2}^{\beta/2}d\tau\bar{\eta}_m\left(D_m+i\dot{\theta}_m\sigma^x\right)\eta_m\right\},\label{appcor1}
\end{split}
\end{align}
where 
\begin{equation}
\Sigma\equiv\sigma^z-i\sigma^y=\left(\begin{array}{cc} 1 & -1 \\1 & -1 \end{array} \right).
\end{equation}
To calculate the expectation value that appears in (\ref{appcor1}) we introduce the generating functional
\begin{equation}
Z_m[g]=\int_{AP}\mathcal{D}\bar{c}_m\mathcal{D}c_m\mathcal{D}{\bar{c}}_{N-m-1}\mathcal{D}c_{N-m-1}e^{-S_m[g]}, \label{appgen1}
\end{equation}
where the action entering in the last integral assumes the form
\begin{equation}
S_m[g]=\int_{-\beta/2}^{\beta/2}d\tau\left(\bar{\eta}_m\tilde{D}_m\eta_{m}+g(\tau)\bar{\eta}_m\Sigma\eta_m\right). \label{appact1}
\end{equation} 
After the calculation of the path integral (\ref{appgen1}) the correlation function (\ref{appcor1}) will be found by taking the functional derivative of the result as
\begin{equation}
B_{ba}^{(eq)}(\tau)=-\frac{1}{N}\sum_{m=0}^{N-1}e^{\frac{2\pi{i}}{N}(m+\frac{1}{2})(b-a)-2i\theta_m}\frac{\delta\text{ln}Z_m[g]}{\delta{g}(\tau)}\bigg|_{g=0}. \label{deriv}
\end{equation}
After these abbreviations, the generating functional in (\ref{appgen1}) can be recasted into the following compact expression
\begin{equation}
Z_m[g]=\text{Det}(\tilde{D}_m)\braket{\text{exp}\left(-\int_{-\beta/2}^{\beta/2}d\tau{g(\tau)}\bar{\eta}_m\Sigma\eta_m\right)}, \label{appgenexp1}
\end{equation}
where the definition of the expectation value appearing in the last expression is as follows
\begin{equation}
\braket{\left(...\right)}\equiv\left[\text{Det}\left(\tilde{D}_m\right)\right]^{-1}\int_{AP}\mathcal{D}\bar{c}_m\mathcal{D}c_m\mathcal{D}{\bar{c}}_{N_m-1}\mathcal{D}c_{N-m-1}e^{-\int_{-\beta/2}^{\beta/2}d\tau\bar{\eta}_m\tilde{D}_m\eta_m}\left(...\right). \label{appexp1}
\end{equation}
To continue we can use the Cluster Expansion theorem \cite{33} to express the expectation value in (\ref{appgenexp1}) in terms of the connected correlation functions as
\begin{equation}
\braket{\text{exp}\left(-\int_{-\beta/2}^{\beta/2}d\tau{}M_m(\tau)\right)}=\text{exp}\left[\sum_{n=1}^{\infty}\frac{(-1)^n}{n!}\int_{-\beta/2}^{\beta/2}d\tau_1...\int_{-\beta/2}^{\beta/2}d\tau_n\braket{M_m(\tau_1)...M_m(\tau_n)}_c\right]. \label{clust1}
\end{equation}
For the case in hand we only need the first term in the exponential factor of the last expression, which contains only a single $g(\tau)$ function:
\begin{equation}
Z_m[g]=\text{Det}\left(\tilde{D}_m\right)\text{exp}\left(-\int_{-\beta/2}^{\beta/2}d\tau{g(\tau)\braket{\bar{\eta}_m\Sigma\eta_m}_c}+\mathcal{O}(g^2)\right).\label{result}
\end{equation}
Combining Eqs. (\ref{deriv}) and (\ref{result}) we find
\begin{equation}
B_{ba}^{(eq)}(\tau)=-\frac{1}{N}\sum_{m=0}^{N-1}e^{\frac{2\pi{i}}{N}\left(m+\frac{1}{2}\right)(b-a)-2i\theta_m(\tau)}\text{tr}\braket{\tau|\Sigma\tilde{G}_m|\tau},\label{coor}
\end{equation}
where $\tilde{G}_m$ the Green's function satisfying $(D_m+i\dot{\theta}_m\sigma^x)\tilde{G}_m=\mathbb{1}$. For the calculation of $\tilde{G}_m$ we rewrite it in the form
\begin{equation}
\tilde{G}_m=\frac{\mathbb{1}}{\mathbb{1}+K_m}G_m.\label{green}
\end{equation}
By applying the Helmholtz technique \cite{34} we can interpret (\ref{green}) in the following form
\begin{equation}
	\tilde{G}_m=\left(\sum_{n=0}^{\infty}\frac{(-1)^{n}}{n!}a^{(m)}_n\right)^{-1}\left(\sum_{k=0}^{\infty}\frac{(-1)^{k}}{k!}b^{(m)}_k\right)G_m, \label{beforehelm}
\end{equation}
where 
\begin{equation}
a^{(m)}_n=-\text{det}\left(\begin{array}{cccccc} 1 & 0 & 0 & . & 0 & \text{Tr}K_m \\\text{Tr}K_m & 2 & 0 & . & 0 & \text{Tr}K_m^2\\
\text{Tr}K^2_m & \text{Tr}K_m & 3 & . & 0 & \text{Tr}K_m^3 \\
. & . & . & . & . & .\\
. & . & . & . & . & .\\
\text{Tr}K^{n-1}_m & \text{Tr}K^{n-2}_m & . & . & \text{Tr}K_m & \text{Tr}K^n_m\end{array} \right),\quad{n\geq2} 
\end{equation}
with $a^{(m)}_0=1$, $a^{(m)}_1=-\text{Tr}K_m$ and each $b^{(m)}_k$ is found by substituting $\text{Tr}K^n_m\rightarrow\text{Tr}K_m^n-K_m^n$ in $a^{(m)}_k$.
Thus, each order can be systematically understood from this expansion.
To define this as a single sum, we define as $A^{(m)}_n$ the term involving all terms of the form $K^k_m\text{Tr}K_m^l$ for which $k+l=n$. Thus, we can symbolically write the expansion $\tilde{G}_m=\sum_{n=0}^{\infty}A^{(m)}_nG_m$ which in turn gives for the correlator (\ref{coor}):
\begin{equation}
\text{tr}\braket{\tau|\Sigma\tilde{G}_m|\tau}=1+\sum_{\nu=1}^{\infty}\text{tr}\braket{\tau|\Sigma{{A}_m^{(\nu)}{G}_m}|\tau}\equiv1+\sum_{\nu=1}^{\infty}c^{(\nu)}_m(\tau). \label{ser}
\end{equation}
To analyse the asymptotic behaviour of this function it is enough to consider the first term in the expansion, coming from ${A}^{(1)}_m=-K_m$, which after rescaling $\omega\tau=\sigma$ and taking the limit $\beta\rightarrow\infty$ is readily seen to have the form:
\begin{align}
\begin{split}
c_m^{(1)}(\tau)&=-i\text{tr}\braket{\tau|\Sigma{G}_m\dot{\theta}\sigma^xG_m|\tau}=\\
&=i\int_{-\infty}^{\infty}d\sigma_1\dot{\theta}_m(\sigma_1)\left[\Theta(\sigma-\sigma_1)e^{-\frac{2}{\omega}\int_{\sigma_1}^{\sigma}d\tilde{\sigma}\epsilon_m(\tilde{\sigma})}-\Theta(\sigma_1-\sigma)e^{-\frac{2}{\omega}\int_{\sigma}^{\sigma_1}d\tilde{\sigma}\epsilon_m(\tilde{\sigma})}\right],
\end{split}
\end{align}
where $\sigma=\omega\tau$.
In the adiabatic regime, the transformation (\ref{id}) can again be used, giving for a general magnetic field $h(\tau)$:
\begin{equation}
c^{(1)}_m(\tau)\underset{\omega\rightarrow0,\beta\rightarrow\infty}{\simeq}-i\frac{\omega^2}{2}\frac{1}{\epsilon_m(\sigma)}\partial_\sigma\left(\frac{\dot{\theta}(\sigma)}{\epsilon_m(\sigma)}\right),\quad{\sigma=\omega\tau}.\label{resu1}
\end{equation}
To see how this function behaves as the system approaches the critical point, we again study the case of $h(\tau)=\omega\tau$. The function (\ref{resu1}) is bounded, $|c^{(1)}_m(\tau)|\leq\pi/2$, and satisfies this bound at the sudden limit $\omega\rightarrow\infty$ and at the vicinity of the critical point $|\sigma-1|\sim|\phi|\sim0$:
\begin{equation}
|c^{(1)}_m(\sigma)|\underset{\omega\rightarrow\infty}{\simeq}|2\theta(\sigma)|\underset{\sigma\rightarrow1\pm0}{\leq}\pi/2+\mathcal{O}(1/\omega).\label{last?}
\end{equation}
Substituting the linear magnetic field in Eq. (\ref{resu1}) we find at the adiabatic limit $\omega\rightarrow0$:
\begin{equation}
c_m^{(1)}(\sigma)\underset{\omega\rightarrow0,\beta\rightarrow\infty}{\simeq}i\omega^2\frac{r\text{sin}\phi_m}{\epsilon_m(\sigma)}\partial_{\sigma}\epsilon_m^{-3}(\sigma)+\mathcal{O}(\omega^3).
\end{equation}
Due to the multiple exponential factors, the last result presents the leading behaviour of the series in Eq. (\ref{ser}) as long as the magnetic field is away from its critical value. However, this argument fails at the critical vicinity $|\sigma-1|\sim|\phi_m|\sim\epsilon_m\sim\mathcal{O}(\omega^{1/2})$ where each term in the expansion (\ref{ser}) is $\mathcal{O}(1)$. To understand this behaviour one has to take into account that at the thermodynamic limit the summation over $m$ becomes an integral over the angle $\phi$ as noted in Eq. (\ref{bq}) of the main text. For our purposes it is enough to examine only the contribution of the first term in the expansion (\ref{ser}) where we use the definition of $B^{Q}_l(\sigma)$ given in (\ref{comp}):
\begin{align}
\begin{split}
B_l^{Q(1)}(\sigma)&\underset{N\rightarrow\infty}{=}-\frac{1}{2\pi}\int_{0}^{2\pi}d\phi{e}^{i\phi{l}-2i\theta(\phi;\sigma)}c^{(1)}(\phi;\sigma)=\\
&\underset{\beta\rightarrow\infty}{=}\frac{r}{4\pi{i}}\int_{-\infty}^{\infty}d\sigma_1\int_{0}^{2\pi}d\phi\frac{\text{sin}\phi{e}^{i\phi{l}-2i\theta(\phi;\sigma)}}{(\sigma_1-\text{cos}\phi)^2+\left(r\text{sin}\phi\right)^2}\times\\
&\times\left[\Theta(\sigma-\sigma_1)e^{-\frac{2}{\omega}\int_{\sigma_1}^{\sigma}d\tilde{\sigma}\epsilon_m(\tilde{\sigma})}-\Theta(\sigma_1-\sigma)e^{-\frac{2}{\omega}\int_{\sigma}^{\sigma_1}d\tilde{\sigma}\epsilon_m(\tilde{\sigma})}\right].\label{laastlabel}
\end{split}
\end{align} 
Poles appear when $\sigma_1=1\pm\mathcal{O}(\omega^{1/2})$. However, their contribution is suppressed due to the exponentials in the last factor, except when $\sigma$ approaches its critical value $\sigma=1\pm\mathcal{O}(\omega^{1/2})$. In this critical region the exponential terms reduce to a constant and the remaining integrals can be easily performed leading to the conclusion that $c^{(1)}(\phi;\sigma)$ in the critical vicinity $|\sigma-1|\sim|\phi|\sim\mathcal{O}(\omega^{1/2})$ is almost constant $|c^{(1)}|\simeq\pi/2$. Thus,
\begin{equation}
B_l^{Q(1)}(1)\simeq\frac{i}{4}\int_{-\phi_0}^{\phi_0}d\phi{e}^{i\phi{l}-2i\theta(\phi;1)}=\frac{\text{sin}[\phi_0(2l+1)/2]}{2l+1},\quad\phi_0=\mathcal{O}(\omega^{1/2}).
\end{equation}
This argumentation can be extended to the full term $B_l^{Q}$ in Eq. (\ref{ser}), justifying the conclusion that at the vicinity of the critical point $c^{(\nu)}\simeq\mathcal{O}(1)$. At the critical point $\sigma=1$ and at the sudden limit $\omega\rightarrow\infty$ the exponential factors in Eq. (\ref{laastlabel}) can also be neglected leading to a result similar to the one obtained at the adiabatic limit (see Eq. (\ref{last?})).

\end{document}